\documentclass[twocolumn,preprintnumbers]{revtex4-1}
\usepackage{amsmath}
\usepackage{amssymb} 
\usepackage{color}
\usepackage{float}
\usepackage{sistyle}
\usepackage{subfigure}  
\usepackage{verbatim}   
\usepackage{graphicx}
\usepackage{dcolumn}
\usepackage{bm}
\usepackage[citecolor=blue,filecolor=blue,colorlinks=true,urlcolor=blue]{hyperref}
\usepackage{ulem}
\usepackage{lineno}

\newcommand{\BAN}{\ensuremath{B_{1g}\,}}
\newcommand{\BN}{\ensuremath{B_{2g}\,}}

\newcommand{\Ts}{\ensuremath{T^{\ast}\,}}
\newcommand{\Tc}{\ensuremath{T_{\rm c}\,}}
\newcommand{\cm}{\ensuremath{{\rm cm}^{-1}}}

\graphicspath{{images/}{../images/}}

\begin{document}

\title{Intimate link between Charge Density Wave, Pseudogap and Superconducting Energy Scales in Cuprates}

\author{B. Loret$^1$, Y. Gallais$^1$, M. Cazayous$^1$, A. Forget$^2$, D. Colson$^2$, M.-H. Julien $^4$,
I. Paul$^1$, M. Civelli$^3$, A. Sacuto$^1$${^\ast}$}

\affiliation{$^1$ Laboratoire Mat\'eriaux et Ph\'enom$\grave{e}$nes Quantiques (UMR 7162 CNRS), Universit\'e Paris Diderot-Paris 7, Bat.Condorcet, 75205 Paris Cedex 13, France,\\
$^2$ Service de Physique de l'{\'E}tat Condens{\'e}, DSM/IRAMIS/SPEC (UMR 3680 CNRS), CEA Saclay 91191 Gif sur Yvette cedex France.\\
$^3$ Laboratoire de Physique des Solides, CNRS, Univ. Paris-Sud, Universit\'e Paris-Saclay, 91405 Orsay Cedex, France\\
$^4$ Laboratoire National des Champs Magn\'etiques Intenses, CNRS-Universit\'e Grenoble Alpes-Universit\'e Paul Sabatier-Institut National des Sciences Appliqu\'ees, European Magnetic Field Laboratory, 38042 Grenoble, France}

\date{\today}



\maketitle

\textbf{
The cuprate high temperature superconductors develop spontaneous charge density wave (CDW)
order below a temperature $T_{\rm CDW}$ and over a wide range of hole doping ($p$).
An outstanding challenge in the field is to understand whether this modulated
phase is related to the more exhaustively studied pseudogap and superconducting phases
\cite{Fradkin:2014,Keimer2015}. 
To address this issue it is important to extract
the energy scale $\Delta_{\rm CDW}$ associated with the charge modulations, and to compare it with the pseudogap (PG)
$\Delta_{\rm PG}$ and the superconducting gap $\Delta_{\rm SC}$. However, while $T_{\rm CDW}$ is well-characterized from
earlier works \cite{Comin2016} little has been known about $\Delta_{\rm CDW}$ until now. Here, we report the extraction of $\Delta_{\rm CDW}$
for several cuprates using electronic Raman spectroscopy. Crucially, we find that, upon approaching the parent Mott state
by lowering $p$, $\Delta_{\rm CDW}$ increases in a manner similar to the doping dependence of $\Delta_{\rm PG}$ and $\Delta_{\rm SC}$.
This shows that CDW is an unconventional order, and that the above three phases are controlled by the same electronic correlations.
In addition, we find that $\Delta_{\rm CDW} \approx \Delta_{\rm SC}$ over a substantial doping range,
which is suggestive of an approximate emergent symmetry connecting the charge  modulated phase with superconductivity
\cite{Efetov13,Sachdev13,Davis2013,Hayward2014,Wang2015,Montiel2017}.}\\

In recent years, many experiments and different techniques have established the ubiquity of CDW order in cuprates \cite{Comin2016}.
In particular, these works have determined $T_{\rm CDW}(p)$, which displays a dome-like shape on the temperature-doping ($T-p$) phase diagram, in a fashion reminiscent of the superconducting dome $T_{\rm SC}(p)$, even though the former order is present over a much narrower $p$-range, and mostly below optimal doping. The CDW is found to compete with superconductivity \cite{Tranquada95,Hoffman02,Wu11,Ghiringhelli12,Chang12,Leboeuf13,Taillefer2015} but there are indications that the interplay between the two phenomena might be more complex than a simple competition \cite{Kacmarcik2018,Edkins2018}.

The energy scale $\Delta_{\rm CDW}$ associated with the CDW has attracted far less experimental attention, even though
this quantity is crucial to address several important questions such as the following.
(a) First, whether the CDW is a conventional order i.e., a phase whose existence can be understood within a scenario of weakly interacting
electrons. A tell-tale signature of it would be if $T_{\rm CDW}(p) \propto \Delta_{\rm CDW} (p)$.
On the other hand if their doping trends are different, as is famously the case of the superconducting order,
it implies unconventional order, which is a consequence of strongly interacting electrons. 
Here we show that this is also the case of the CDW and, therefore, it is an unconventional order.
(b) Second, a comparison of the magnitudes and the doping dependencies of $\Delta_{\rm CDW} (p)$, $\Delta_{\rm SC}(p)$ and
$\Delta_{\rm PG}(p)$ is important to understand the relation between these three phenomena. We show that these
three energy scales have rather similar doping evolutions, implying that it is likely that they have a common origin
in terms of a driving electronic interaction. Moreover, we find that the magnitude of $\Delta_{\rm CDW} (p)$ and
of $\Delta_{\rm SC}(p)$ are comparable over a significant doping range, which is consistent with a concept that has
gained importance in recent times, namely the presence of an emergent approximate symmetry that links CDW, which is a particle-hole instability, with superconductivity, which is an instability involving particle-particle excitations. This symmetry is exact only at the so called ``hot spots'' of the Fermi surface \cite{Efetov13,Sachdev13,Davis2013,Hayward2014,Wang2015,Montiel2017}.

\begin{figure}[h]
\begin{center}
\includegraphics[scale=0.6]{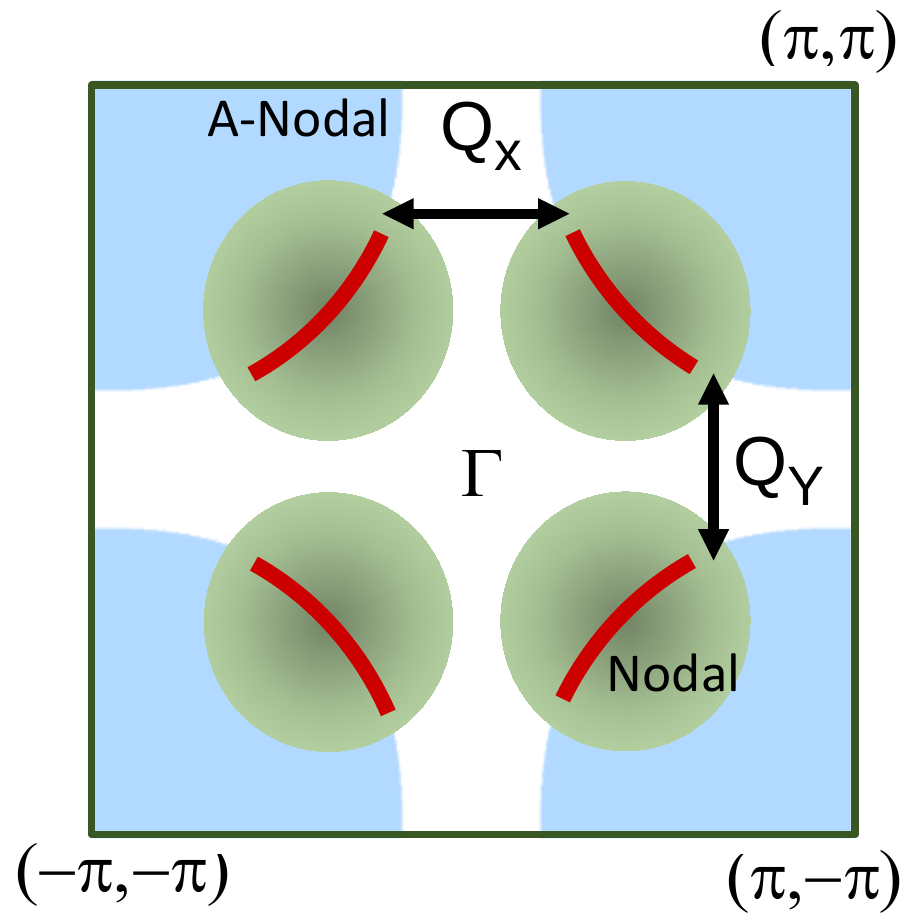}
\caption{(Color online). The white region represents the tight-binding Fermi volume of hole-doped cuprates 
in the first Brillouin zone calculated from \cite{Tanmoy2012}, (see SI). Red arcs indicate the actual experimental Fermi surface observed e.g. by photo-emission \cite{Keimer2015}. The $Q_x$ and $Q_y$ ordering vectors reproduce the bi-collinear CDW, as observed e.g. with X-rays \cite{Comin2016}. Green zones highlight the nodal regions that are probed in the \BN Raman response.}
\label{fig:1}
\end{center}\vspace{-7mm}
\end{figure}

A typical signature of a density wave in Raman spectroscopy is the loss of spectral weight of the electronic continuum at low energy, followed by a recovery of spectral weight at higher energy, as the order sets in as a function of temperature \cite{Ralevic2016}. 
In the cuprates, however, since $T_{\rm CDW} < T^{\ast}$, the characteristic PG temperature, one technical challenge is to distinguish CDW from the loss of spectral weight due to the PG itself \cite{Sacuto2013}. As we show below, this can be overcome by studying the \BN Raman response, which preferentially probes the nodal regions of the Brillouin zone (Fig. \ref{fig:1}), and where PG effects are known to be minimal \cite{Norman1998}. This intuition is further aided by the fact that the Bloch states that are  expected to reconstruct the most due to the CDW are in between the nodal and the anti-nodal regions, as indicated by the CDW ordering vectors $Q_x$ and $Q_y$ on Fig. \ref{fig:1}. \cite{Sebastian2015,Comin2016}.  Therefore, we anticipate that the \BN Raman geometry should be most favorable to search for a signature of the CDW.

A second technical challenge, from a materials point of view, is to identify the cuprates family which has
the cleanest $CuO_2$ layer where the CDW order can set in.
Here, we identified the inner $CuO_2$ layer of  HgBa$_2$Ca$_2$Cu$_3$O$_{8+\delta}$ (Hg-1223)
 to be most suitable because it is homogeneously doped and screened from out-of-plane disorder by the outer planes,
as demonstrated by the analysis of the $^{63}$Cu-NMR line-width \cite{Julien1996, Mukuda2012}.

\begin{figure*}[!ht]
\begin{center}
\includegraphics[scale=0.2]{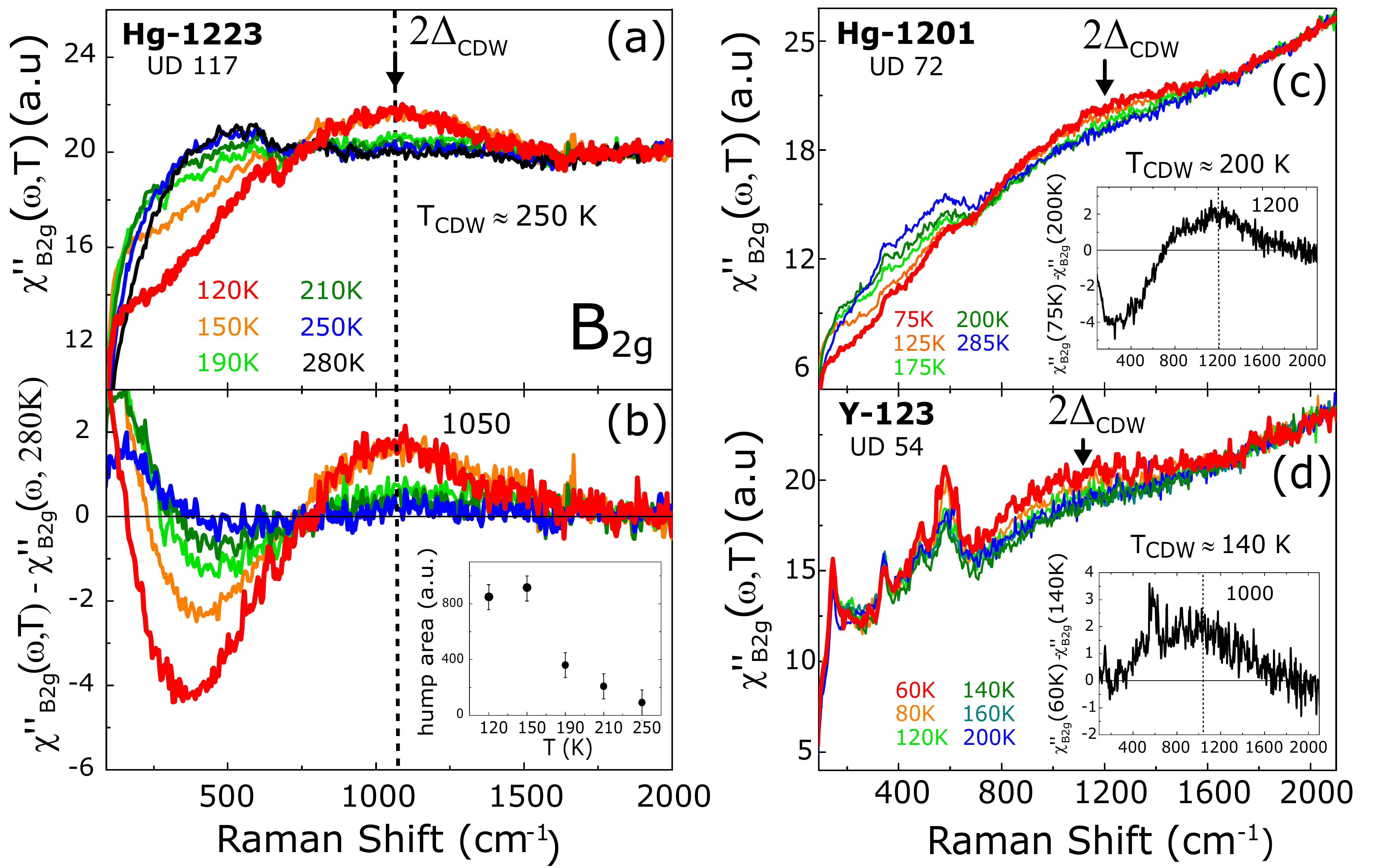}
\caption{(Color online).  (a) Temperature dependence of the \BN (nodal) Raman responses of  
HgBa$_2$Ca$_2$Cu$_3$O$_{8+\delta}$ compound above \Tc = 117 K. (b) The same responses, after subtracting the one at 280 K to highlight the CDW gap structure (dip and hump). In the inset, we show the CDW hump area as function of temperature.  
Temperature dependence of the \BN Raman response of (c) HgBa$_2$CuO$_{4+\delta}$ (Hg-1201) with \Tc=72 K and of (d) YBa$_2$Cu$_3$0$_{7-\delta}$ (Y-123) with \Tc= 54 K. In the insets, we display the difference between the Raman responses measured at \Tc and the the ones at $T_{\rm CDW}$.} 
\label{fig:2}
\end{center}\vspace{-7mm}
\end{figure*}

We performed electronic Raman scattering measurements on Hg-1223 single crystals grown with a single step synthesis \cite{Loret2017}.
Given the positive results in Hg-1223 (see below), we also measured single crystals from the HgBa$_2$CuO$_{4+\delta}$ (Hg-1201) and the YBa$_2$Cu$_3$0$_{7-\delta}$ (Y-123) families in order to demonstrate that the CDW signature in Raman response is present in several types of cuprates. The experimental details are given in the Supplementary Information (SI).

Our first central observation is that the \BN Raman response $\chi^{\prime \prime}_{\BN} (\omega, T)$ of an underdoped Hg-1223 crystal displays a well defined gap-structure, typical of a density wave, at 120 K, which is above \Tc = 117 K but below the pseudogap temperature \Ts $\approx$ 360 K as defined from the NMR Knight shift at this doping level \cite{Julien1996} (Fig.2 (a)). 
It is characterized by a hump in the electronic background centered around $2 \Delta_{\rm CDW}$ = 1050 \cm, which we take as the energy scale associated with the CDW order. It is accompanied below 750 \cm  by a depletion with respect to the background measured at 280 K, as better shown in Fig.2 (b) by $\chi^{\prime \prime}_{\BN} (\omega, T)- \chi^{\prime \prime}_{\BN} (\omega, 280$ K). As the temperature increases, the low energy electronic depletion fills up while the hump at $2 \Delta_{\rm CDW}$ decreases in intensity, until it disappears at $T_{\rm CDW} \approx 250$ K, as shown by the integrated hump area in the inset. Note that, since the PG signature is known to be very weak in the \BN Raman response, this allows us to observe the CDW gap. The \BAN Raman response, on the other hand, shows clear loss of spectral weight related to the PG which starts above 280 K (see Fig. \ref{fig:S2}(d) in the SI).

Furthermore, we find that a very similar CDW gap structure is also observed in the \BN Raman response of under-doped Hg-1201 and Y-123 cuprates, which have one and two CuO$_2$ layers, respectively (Fig. \ref{fig:2}(c, d)). The CDW hump is located around $2\Delta_{\rm CDW} \simeq 1200 \, \cm$ for the Hg-1201 crystal (\Tc =72 K, p=0.09) while it is around $1000 \,\cm$ in the case of Y-123 crystal (\Tc= 54 K, p=0.10). These features disappear around $T_{\rm CDW}$ $\approx$ 200 K and 140 K, respectively. In the under-doped Y-123 compound the low frequency loss of spectral weight is less pronounced, possibly masked by the phonon peaks at those frequencies, or because of oxygen disorder in the Cu-O chain and of twinned crystals. Nevertheless, the results show that the CDW gap structure is a common feature of the \BN responses of the under-doped cuprates.

Next we show theoretically that the formation of the CDW leads to a $B_{2g}$ signal with
low-frequency depletion followed by a hump at a higher frequency, which is qualitatively in line with what is
reported in Fig. 2. We consider the simplest tight-binding model of the cuprates
$\mathcal{H}_0= \sum_{\mathbf{k}} \xi_{\mathbf{k}} \, c^{\dagger}_{\mathbf{k}} c^{}_{\mathbf{k}}$, where
$\xi_{\mathbf{k}}$ is the electron energy dispersion (whose  Fermi surface is displayed in Fig.\ref{fig:1}) 
on the two dimensional square lattice that describes the Cu-O planes (see SM and Ref.\cite{Tanmoy2012}). 
$\mathbf{k}$ is the Bloch wave-vector in the reciprocal space. 
To this we add the CDW potential \cite{Allais14c}
$\mathcal{H}_{CDW} = \, \sum_{\mathbf{k}, \mathbf{Q}}\, f_{\mathbf{Q}}(\mathbf{k}) c^{\dagger}_{\mathbf{k}+\mathbf{Q}/2}
c^{}_{\mathbf{k}-\mathbf{Q}/2} $, with 
$f_{\mathbf{Q}}(\mathbf{k}) = V_{CDW} \left( \cos k_x - \cos k_y \right)$.
We suppress the spin index since it does not play in role in the following.
$\mathbf{Q}_x= (\pi/2, 0)$ and $\mathbf{Q}_y= (0, \pi/2)$ describe a bi-collinear charge modulation of four unit-cell 
periodicity in the $x$ and $y$ directions \cite{Comin2016}.
We ignore that in the real systems the order may be incommensurate, since we do not expect that this simplification 
will affect the result qualitatively. Furthermore, since we are mostly interested in the $B_{2g}$ response 
where the PG is weak, we do not include it in the calculation. It is useful to keep track of the modifications of the dispersion brought about by the CDW potential.
\begin{figure}[!h]
\begin{center}
\includegraphics[scale=0.35]{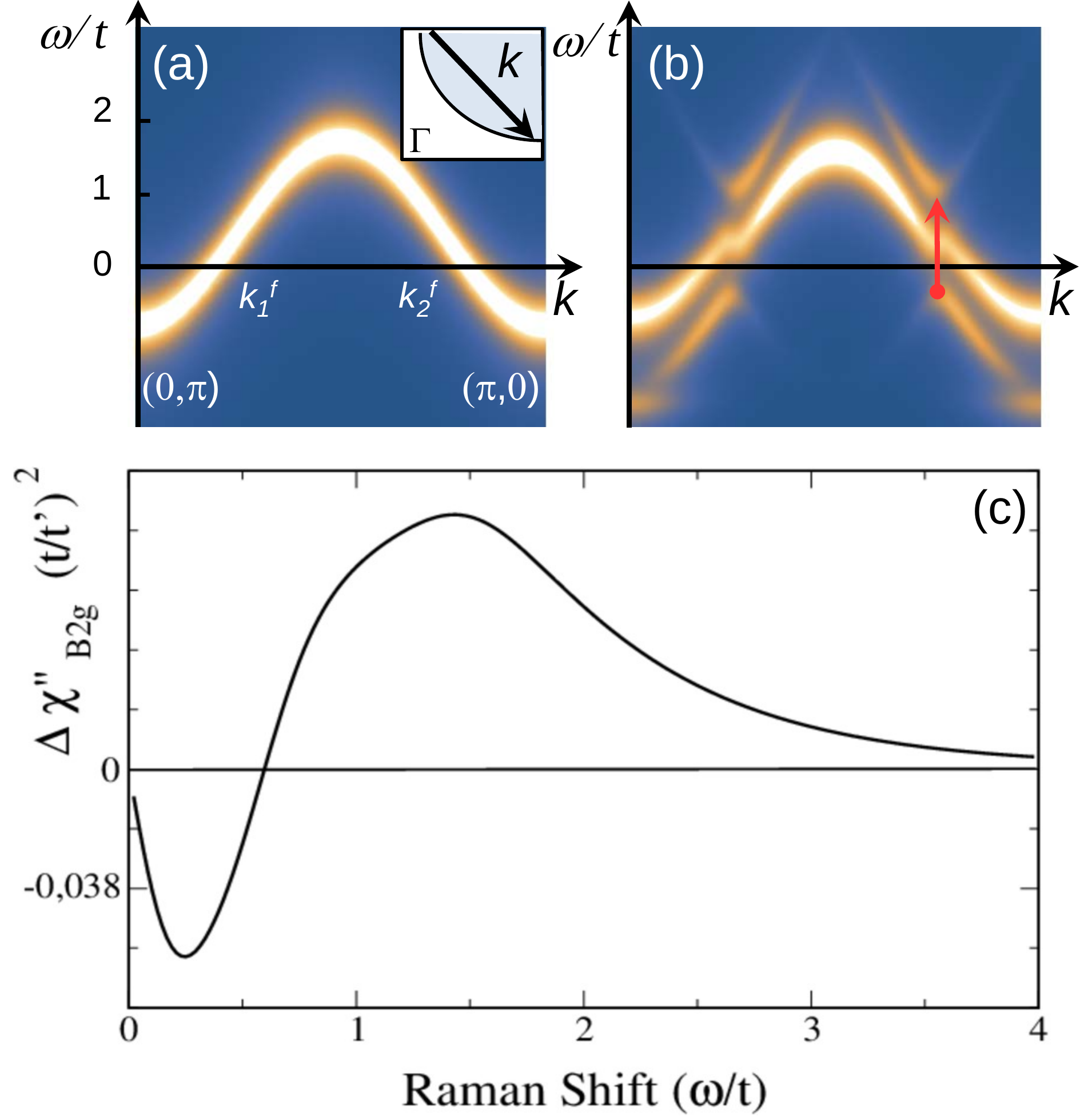}
\caption{(Color online). Theoretical spectral density without (a) and with (b) CDW order
along the $(0,\pi)\to (\pi,0)$ path in the first quadrant of the Brillouin Zone. (c) Difference between the theoretical \BN Raman responses with and without the CDW potential which highlights the dip-hump CDW feature.}
\label{fig:3}
\end{center}\vspace{-7mm}
\end{figure}
In Fig. 3 (a) and (b) we plot the spectral density, $\mathcal{A}(\mathbf{k},\omega)$ (see SI), along the path
$(0,\pi) \to (\pi,0)$ for $V_{CDW}=0$ and $V_{CDW} \neq 0$, respectively.
In the first case the band crosses the Fermi level ($\omega=0$) 
at two Fermi points $k^{f}_{1}$ and $k^{f}_{2}$.
However, in the CDW phase a wide spectral weight re-distribution takes place: even if full gaps do not open at the Fermi level \cite{Verret2017}, cone-like bands appear above and under the Fermi level. This is the effect of the folding of the original bands  into the 16-time reduced Brillouin zone. Notice that in the unoccupied side ($\omega>0$) close to the nodal region, the cone-like bands have sizable spectral weight (indicated by the tips of the arrow in Fig.3 (b)). Consequently,
we expect that the CDW inter-band transition (marked by the arrow) produces a significant signature in the \BN nodal Raman response at an energy close to the arrow length, $\omega_{\uparrow}/t$.
In Fig. 3 (c) we show the difference between the \BN Raman responses calculated with and without the
CDW potential $\Delta\chi^{\prime \prime}_{\BN}(\omega) \equiv \, \chi^{\prime \prime}_{\BN}(\omega,V_{\rm CDW})- \chi^{\prime \prime}_{\BN}(\omega,0)$. We note the the dip-hump feature in $\Delta\chi^{\prime \prime}_{\BN}(\omega)$, where the position of the hump is $\omega= 2 \Delta_{ \rm CDW}\simeq 1.4 t\simeq \omega_{\uparrow} \propto 2 V_{\rm CDW}$,
in qualitative agreement with Fig. 2(b).

\begin{figure}[!h]
\begin{center}
\includegraphics[scale=0.033]{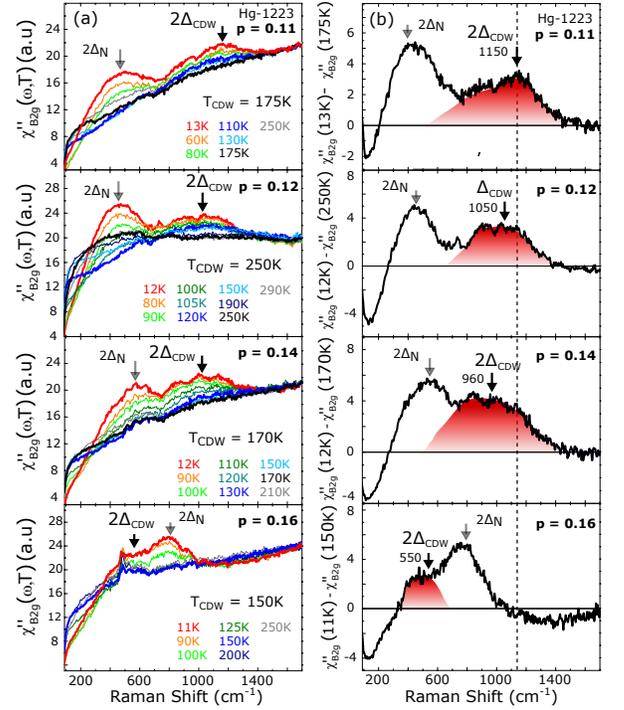}
\caption{(Color online).  Left panels: Temperature dependence of the \BN Raman responses of several
under-doped Hg-1223. \Tc= 105 K  ($p=$ 0.11), \Tc= 117 K ($p=$ 0.12), \Tc= 127 K ($p=$ 0.14), and \Tc = 133 K ($p=$ 0.16). The doping levels were estimated from the empirical Presland-Tallon's law \cite{Presland91}. Right panels: Difference between the SC Raman response at the lowest temperature and the Raman response at $T_{\rm CDW}$. The shaded red form is a guide for the eyes to follow the doping dependence of the CDW energy gap $2 \Delta_{\rm CDW}$.}
\label{fig:4}
\end{center}\vspace{-7mm}
\end{figure}

Our next goal is to map out the doping dependence of $T_{\rm CDW}(p)$ and the energy scale $\Delta_{\rm CDW}(p)$. The left panel of Fig. 4 displays the temperature dependence of the \BN responses of various Hg-1223 samples, from an under-doped ($p=$ 0.11) to an optimally doped ($p=$ 0.16) one. As in Fig. 2, for each compound, $T_{\rm CDW}(p)$ is extracted from the disappearance of the hump. Similar analysis was performed on Raman spectra of the Hg-1201 and Y-123 (shown in Fig. \ref{fig:2} and in the SI, Fig. \ref{fig:S3}). 
Our extracted $T_{\rm CDW}(p)$ matches very well with that obtained with other techniques, 
as reported in Fig. \ref{fig:S1} of the SI. This is additional confirmation that the hump feature highlighted in Fig. \ref{fig:2} is indeed related to the CDW. 

Next we focus on the energy scale $\Delta_{\rm CDW}(p)$. Since an order is fully developed only at zero temperature, ideally $\Delta_{\rm CDW}$ should be determined from the spectra at the lowest available temperature. 
However, in the cuprates, the situation is complicated by the appearance of superconductivity. Nevertheless, while the intensity of the hump has strong T dependence, its position changes little with temperature, therefore we can still keep track the CDW hump in the superconducting phase. Consequently, we can estimate $\Delta_{\rm CDW}$ from the lowest $T$ superconducting spectra without having to weaken superconductivity, e.g. with high magnetic field. Notice that superconductivity gives rise to a $2\Delta_{\rm N}$ peak in the \BN spectra, which has been widely studied in previous works~ \cite{LeTacon06,Devereaux-RMP,Sacuto2013}. We shall not consider it here, but rather focus on the CDW hump, which is better indicated by the shaded red region on $\chi^{\prime \prime}_{\BN} (\omega, T \simeq 10\, K)- \chi^{\prime \prime}_{\BN} (\omega, T_{\rm CDW})$, as displayed in the right side of Fig. 4. The most striking feature is that, with increasing $p$, the  $2\Delta_{\rm CDW}$ energy scale decreases. 
\begin{figure}[!h]
\begin{center}
\includegraphics[scale=0.22]{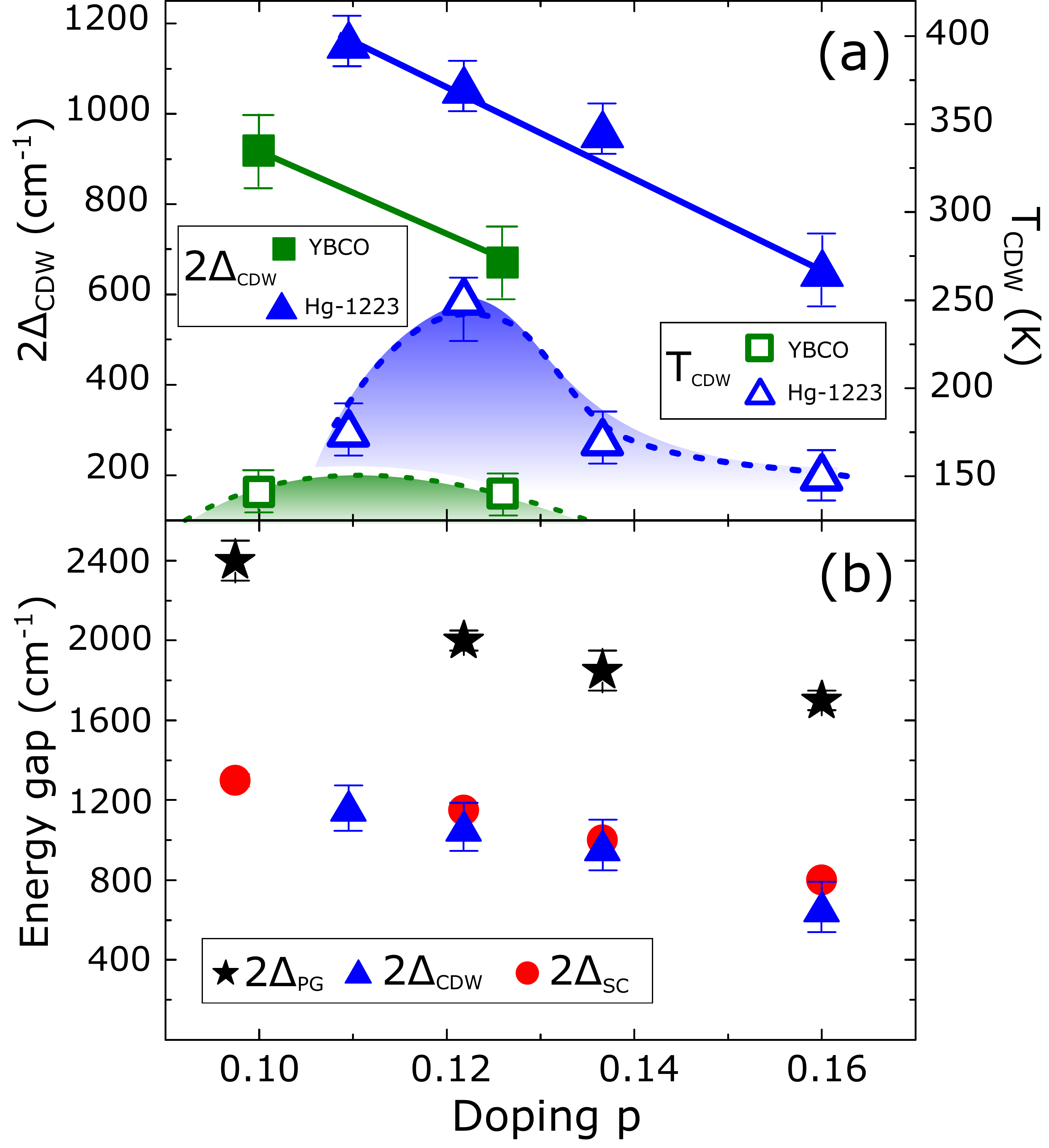}
\caption{(Color online). (a) Doping dependence of $\Delta_{\rm CDW}$ (filled symbols) and $T_{\rm CDW}$ (open symbols) for Hg-1223 (triangles) and Y-123 (squares) cuprates, showing the unconventional character of the CDW phase. Data for Hg-1223 and Y-123 cuprates are extracted from Fig. 4 and Fig. 2 and from the SI (Fig.\ref{fig:S3}). The continuous and doted lines are guide for the eyes; (b) $\Delta_{\rm CDW}$, $\Delta_{\rm SC}$ and $\Delta_{\rm PG}$ display the same doping dependency,
in particular $\Delta_{\rm SC}$ and $\Delta_{\rm CDW}$ are close in energy.} 
\label{fig:5}
\end{center}\vspace{-7mm}
\end{figure}
In fact, we immediately note that $\Delta_{\rm CDW}(p)$ is monotonic and does not follow the dome-like shape of $T_{\rm CDW} (p)$ (see Fig. \ref{fig:5}(a)). 
We also report a similar analysis for the Y-123 compound (see Fig. \ref{fig:2} and Fig. {fig:S3} of SI). 
In other words, the transition temperature and the energy scale associated with the CDW are not proportional to each other. This is reminiscent of the dichotomy between the dome-like behavior of the superconducting $\Tc (p)$ and the anti-nodal superconducting gap $\Delta_{\rm SC}(p)$, which is a hallmark of an unconventional order, i.e. an instability that cannot be understood within scenarios of weakly interacting electrons. 
In Fig. \ref{fig:5}(b) we finally compare the doping dependency of the energy scales $2\Delta_{\rm CDW}(p)$ with the pseudogap $2\Delta_{\rm PG}(p)$ and the superconducting $2 \Delta_{\rm SC}(p)$. This latter is measured from the antinodal \BAN response and  gives direct access to the pairing energy scale, contrary to the nodal Raman $2\Delta_{\rm N}$, which is strongly dependent on the length of the Fermi arcs around the nodes \cite{Sacuto2013}. For the Hg-1223 compound $2\Delta_{\rm SC}(p)$ and $2\Delta_{\rm PG}(p)$ are extracted from the \BAN Raman response reported in Fig. \ref{fig:S2} (a) and (b) of the SI and elsewhere \cite{Loret2016,Loret2017a}. One remarkable point is that $2\Delta_{\rm CDW}(p)$, $2 \Delta_{\rm SC}(p)$ and $2 \Delta_{\rm PG}(p)$ have the same doping dependency which suggests that all these three energy scales are governed by the same electronic interaction. Equally remarkable is the fact that $\Delta_{\rm CDW}(p) \approx \Delta_{\rm SC}(p)$ over a significant doping range. Such near equality of energy scales is an essential perquisite to relate CDW and superconductivity by an emergent approximate symmetry, as it has been proposed in several recent theories \cite{Efetov13,Sachdev13,Davis2013,Fradkin:2014,Hayward2014,Wang2015,Montiel2017}. Consequently, our finding will provide important impetus to theories that propose an intimate link between high temperature superconductivity and CDW in the under-doped cuprates.  
\\
\\

\textbf{Acknowledgements} We acknowledge support from Universit\'e Paris Diderot-Paris 7, CEA, Iramis, SPEC. We thank the Coll\`ege de France and the Canadian Institute for Advanced Research (CIFAR) for their hospitality and support. Correspondence and request for materials should be addressed to A.S. (alain.sacuto@univ-paris-diderot.fr).
\\
\\

\clearpage

\section*{SUPPLEMENTARY INFORMATION}

\subsection{Details of the electronic Raman experiments}

Raman experiments have been carried out using a JY-T64000 spectrometer in single grating configuration using a 600 grooves/mm grating and a  notch filter to block the stray light. The spectrometer is equipped with a nitrogen cooled back illuminated CCD detector. We use the 532 nm excitation line from a diode pump solid state with laser power maintained at 4 mW. Measurements between 10 and 300 K have been performed using an ARS closed-cycle He cryostat. This configuration allows us to cover a wide spectral range ($90~cm^{-1}$ to $2500~cm^{-1}$) with a resolution sets at $5~cm^{-1}$. Spectra have been obtained in one frame. Each frame is repeated twice to eliminate cosmic spikes.  All the spectra have been corrected for the Bose factor and the instrumental spectral response. They are thus proportional to the imaginary part of the Raman response function $\chi^{\prime \prime}(\omega,T)$. The direction of incident electric field is contained in the (ab) plane. The \BN and \BAN geometries are obtained from crossed polarizations, respectively, along and at 45$\arcdeg$ form the Cu-O bond directions. They respectively give access to the nodal region (along the diagonal of the Brillouin zone) and to the anti-nodal region (along the principal axes of the Brillouin zone). The crystal is rotated using a Attocube piezo-rotator ANR 101 put inside the cryostat. 

\subsection{Comparison with previous experiments}

\begin{figure}[!h]
\begin{center}
\includegraphics[scale=0.06]{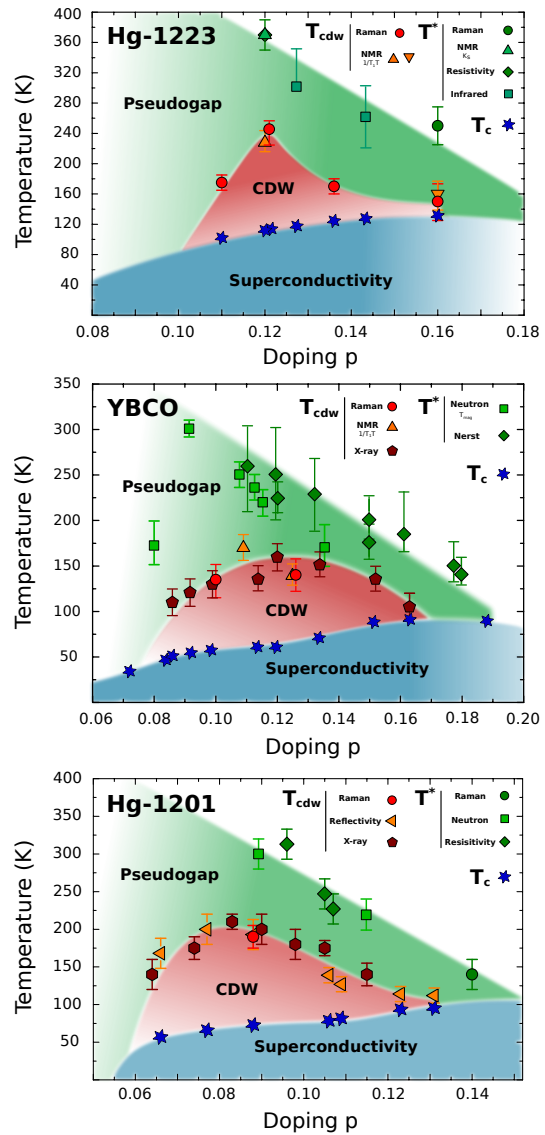}
\caption{(Color online). $T-p$ phase diagrams of (a) Hg-1223, (b) Y-123 and (c) Hg-1201. In panel (a), the $T_{\rm CDW}$ values (extracted from the Hg-1223 Raman spectra of Fig. 4 of the article) are well matched with the NMR data \cite{Julien1996,Mukuda2012}. \Ts values were determined from Raman (present and earlier works \cite{Loret2016,Loret2017a}), NMR \cite{Julien1996}, infrared \cite{McGuire2000}and resistivity \cite{Carrington1994} measurements. In panel (b) the $T_{\rm CDW}$ (extracted from the Y-123 Raman spectra of Fig.2 (d) of the article and Fig.\ref{fig:S3} of the SM) are quite consistent with $T_{\rm CDW} (p)$ obtained from XRD and NMR \cite{Ghiringhelli12, Chang12,Huecker2014,Blanco-Canosa14, Wu2015}. \Ts values were obtained from neutron \cite{Sidis13} and Nernst \cite{Daou2010} measurements. In panel (c), the  $T_{\rm CDW}$ value (extracted from the Hg-1201 Raman spectra of Fig.2 (c) of the article) fits well with the $T_{\rm CDW} (p)$ curve reported by X-ray and transient reflectivity data \cite{Tabis14,Tabis2017,Hinton2016}. \Ts values were obtained from Raman \cite{Guyard2008}, resistivity \cite{Li2008} and neutron measurements \cite{Li2011,Chan2016}.}
\label{fig:S1}
\end{center}\vspace{-7mm}
\end{figure}

The $T_{\rm CDW} (p)$ values extracted from the Raman spectra of Fig.2 and Fig.4 of the article and Fig.\ref{fig:S3} of the SM,  draw a dome like shape in a very similar way than the doping dependence of the onset temperature of the CDW order, $T_{\rm CDW} (p)$, already reported in Hg-1223, Y-123 and Hg-1201 \cite{Hucker13,Blanco-Canosa14,Wu2015,Tabis2017} (see Fig.\ref{fig:S1} (a), (b) and (c)). 

In the absence of X-ray scattering data for Hg-1223, we have used the criterion of maximum of  $1/T1T$ vs $T$ to define the onset of CDW correlations. Although very few NMR studies have been performed on Hg-1223 \cite{Julien1996, Mukuda2012}, the doping dependence of the $1/T1T$ values (orange triangles) fit pretty well with $T_{\rm CDW} (p)$  (red filled circles) as shown in Fig.\ref{fig:S1} (a).  It was recently realized that the maximum in $^{63}$Cu $1/T1T$ curve (where $T1$ is the spin-lattice relaxation time) coincides with $T_{\rm CDW}$ in YBCO \cite{Huecker2014,Wu2015}. This suggests that $1/T1T$ is not related to the pseudogap (as supposed earlier) but rather to the CDW order. 

\subsection{Superconducting gap and Pseudogap in Raman spectroscopy}

\begin{figure}[!h]
\begin{center}
\includegraphics[scale=0.15]{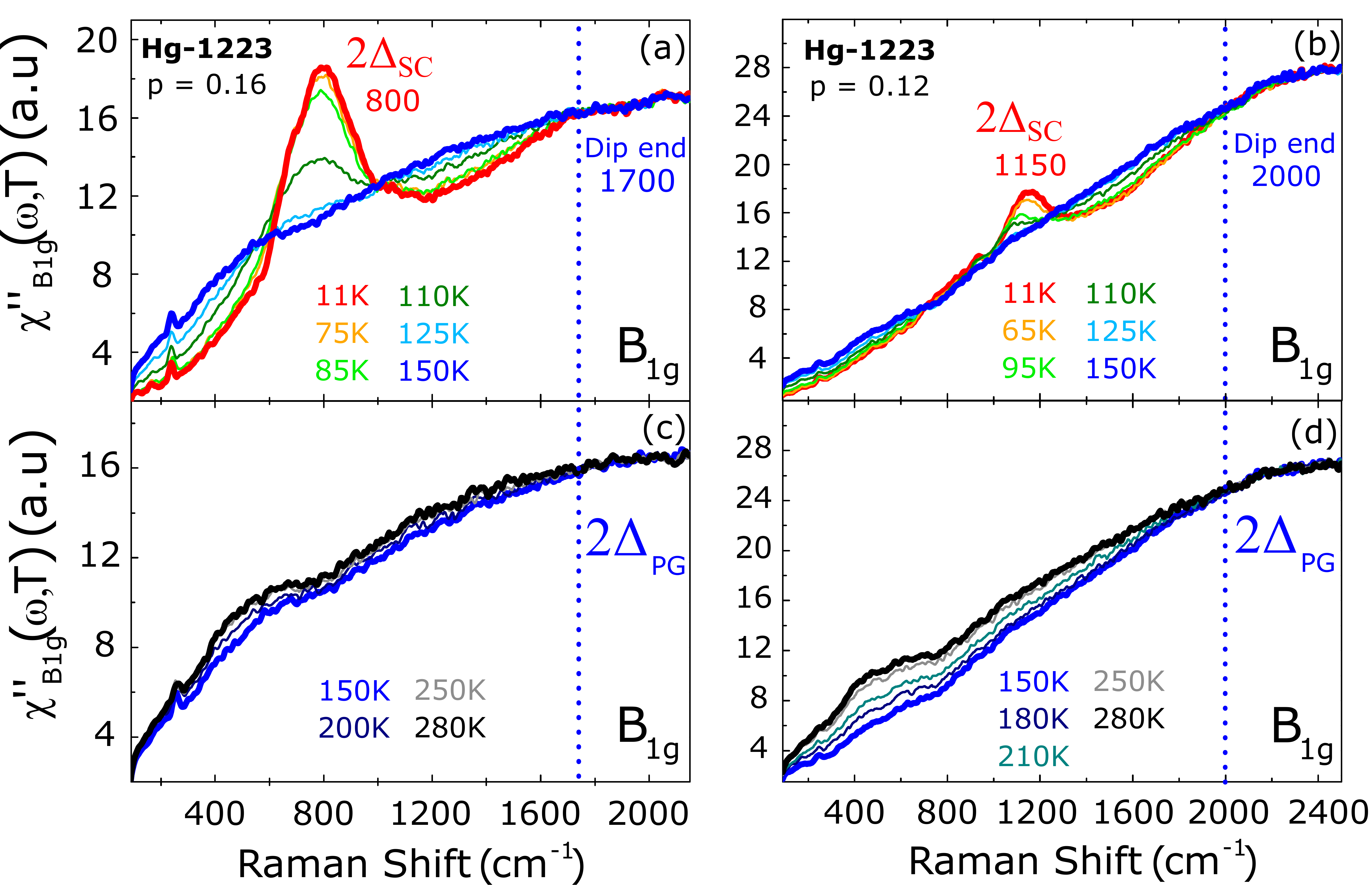}
\caption{Temperature dependence of the \BAN Raman response function of optimally doped Hg-1223 (p=0.16, \Tc=133 K) and under-doped Hg-1223 (p=0.12, \Tc=117K) below $T_c$ (a) and (b) and above $T_c$ (c) and (d) respectively. $2\Delta_{\rm SC}$ and  $2\Delta_{\rm PG}$ are clearly distinct for each doping level.}
\label{fig:S2}
\end{center}\vspace{-5mm}
\end{figure}

We display in Fig.\ref{fig:S2} (a) and (b) the superconducting \BAN Raman responses of Hg-1223 for two distinct doping levels p=0.16 (\Tc=133K) and p=0.12 (\Tc=117K).  They exhibit respectively a pair breaking peak at $2\Delta_{\rm SC} \approx 800$ \cm and $\approx$ 1150 \cm . We can also observe that each $2\Delta_{\rm SC}$ peak is associated on its right energy side to a dip in the electronic continuum. In previous works, we showed this peak-dip structure detected in the superconducting \BAN Raman response results from the interplay between the PG and the SC gap, and can be smoothly connected to the PG appearing in the electronic spectrum above Tc \cite{Loret2016,Loret2017a}. Interestingly, the energy of the dip end detected in the superconducting \BAN Raman response corresponds to the end of the energy range where the loss of spectral weight associated with normal state PG sets in. This is pointed out for the two distinct doping level by dotted line in panels (a),(c) and (b),(d). $2\Delta_{\rm PG}$ is determined from the energy of the dip end or the end of the energy range of the PG depletion. $2\Delta_{\rm PG} \approx$ 1700 \cm for p=0.16 and $2\Delta_{\rm PG} \approx 2000$ \cm for p=0.12.

\subsection{Raman response of underdoped Y-123 (UD 67)}

\begin{figure}[!h]
\begin{center}
\includegraphics[scale=0.4]{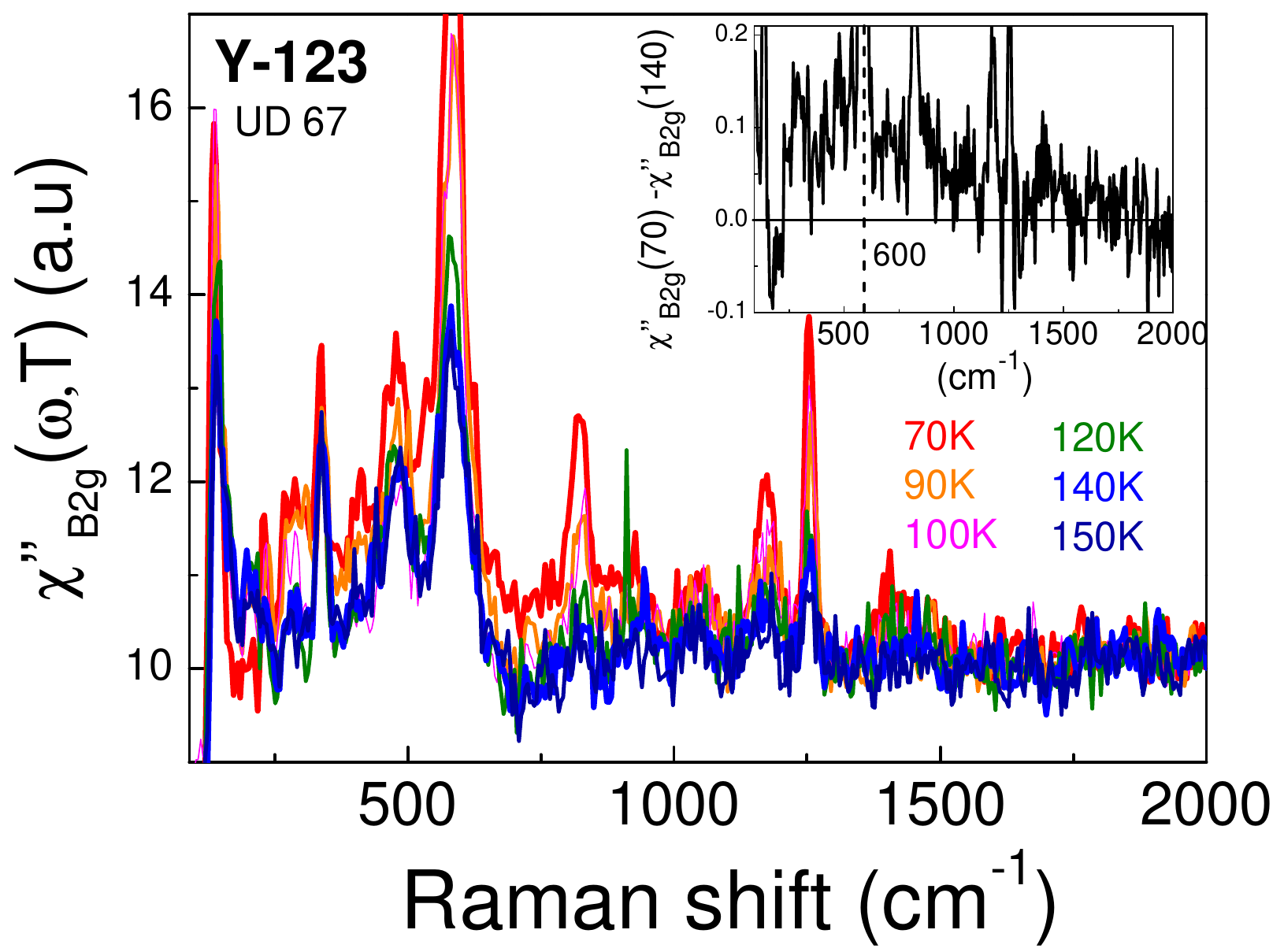}
\caption{Temperature dependence of the \BN Raman response function of the underdoped Y-123 (\Tc=67 K, p=0.13) above $T_c$.  In the inset we display the difference between the Raman responses measured just above \Tc and the the ones at $T_{\rm CDW}$. $\Delta_{\rm CDW}\approx$ 600 \cm. The same linear background has been subtracted from all the Raman responses}
\label{fig:S3}
\end{center}\vspace{-5mm}
\end{figure}

In Fig.\ref{fig:S3} we show the \BN Raman response function of the underdoped Y-123 (\Tc = 67 K, p=0.13). Even if in this compound is certainly more difficult to detect the CDW, possibly because of oxygen disorder in the Cu-O chains and crystals are twinned, the $2\Delta_{\rm CDW}$ is observed as a hump centered around 600 \cm in the Raman spectra (see inset). $T_{\rm CDW} \approx 140$ K is the temperature for which the hump collapses. 

\subsection{Theoretical calculations of the \BN response within a CDW phase}

In order to have a cuprate band structure close to the one that we considered in the experiments, we adopt in $\xi_{\mathbf{k}}$ the band parameters of the Hg-1201 provided by ab-initio calculations\cite{Tanmoy2012}: $t^{\prime}/t= -0.2283$, $t^{''}/t= 0.1739$, $t^{'''}/t= -0.0435$, and we fix the hole doping $p=0.12$, where the CDW phase is strongest on the phase diagram. For the CDW coupling we take $V_{CDW}/t=0.20$, which is a reasonable order of magnitude for these systems\cite{Verret2017}. The bi-collinear charge modulation described by $\mathcal{H}_{CDW}$ describes a four unit-cell periodicity in both the  $x$ and $y$ directions. We can therefore work in the Brillouin zone reduced 16 times by the ordering vectors
$\mathbf{Q}_{\mathbf{n}}= n_1 \mathbf{Q}_x+ n_2 \mathbf{Q}_y$, where $\mathbf{Q}_x= (\pi/2, 0)$ and $\mathbf{Q}_y= (0, \pi/2)$  and $n_1, n_2= 0 \dots 3$. We can rewrite the Hamiltonian: $\mathcal{H}= \, \mathcal{H}_0+ \mathcal{H}_{CDW}$\cite{Verret2017}:

\begin{eqnarray}
\mathcal{H}& =& \, \sum_{\mathbf{k}^{\prime} \in \text{RBZ}}\, \sum_{n_1,n_2= 0\dots 3}\, 
\xi_{\mathbf{k}^{\prime}+ \mathbf{Q}_{\mathbf{n}}}\quad c^{\dagger}_{\mathbf{k}^{\prime}+\mathbf{Q}_{\mathbf{n}}} c^{}_{\mathbf{k}^{\prime}+\mathbf{Q}_{\mathbf{n}}} + 
\nonumber \\
& + & f \left( \mathbf{k}^{\prime}+ \mathbf{Q}_{\mathbf{n}}- \mathbf{Q}_x/2 \right)\,
c^{\dagger}_{\mathbf{k}^{\prime}+\mathbf{Q}_{\mathbf{n}}} c^{}_{\mathbf{k}^{\prime}+\mathbf{Q}_{\mathbf{n}}- \mathbf{Q}_x}+ \nonumber \\
& + & f \left( \mathbf{k}^{\prime}+ \mathbf{Q}_{\mathbf{n}}- \mathbf{Q}_y/2  \right)\,
c^{\dagger}_{\mathbf{k}^{\prime}+\mathbf{Q}_{\mathbf{n}}} c^{}_{\mathbf{k}^{\prime}+\mathbf{Q}_{\mathbf{n}}- \mathbf{Q}_y}+ \nonumber \\
& + & h.c. 
\end{eqnarray}

We can express then the Hamiltonian as a 16$\times$16 matrix $\hat{\mathcal{H}}$ \cite{Verret2017}:
$\mathcal{H}=\, \vec{\psi}^{\dagger}\,(\hat{\mathcal{H}})_{\mathbf{n},\mathbf{n}^{\prime}} \, \vec{\psi}$, where 
we introduce the Nambu notation $\vec{\psi}= ( c_{\mathbf{k}^{\prime}} \dots c_{\mathbf{k}^{\prime}+ n_1 \mathbf{Q}_x + n_2 \mathbf{Q}_y} \dots )$.
The corresponding Green's function is given by $\hat{\mathcal{G}}(\mathbf{k}^{\prime},\omega)=\, (\omega+ \imath \eta- \hat{\mathcal{H}})^{-1}$, where $\eta= 0.20 t$ is a small imaginary part used to display on the real $\omega$ axis. The spectral function can been then obtained 
$\hat{\mathcal{A}}(\mathbf{k}^{\prime},\omega)= -\Im \hat{\mathcal{G}}(\mathbf{k}^{\prime},\omega)/\pi$, and the spectra on the original Brillouin zone   
displayed in Fig.3 (a) and (b) of the article can be derived by unfolding the reduced Brillouin Zone 
$\mathbf{A}(\mathbf{k},\omega)=\, 
\sum_{\mathbf{n}} \, \hat{\mathcal{A}}(\mathbf{k}^{\prime}+ \mathbf{Q}_{\mathbf{n}},\omega)_{\mathbf{n},\mathbf{n}}$.
The Raman \BN response displayed in Fig.3 (c) of the article is calculated (at $T=0$) with the first order bubble approximation:
\begin{eqnarray}
\chi^{''}_{\BN}(\Omega)= \, \sum_{\mathbf{k}^{\prime} \in \text{RBZ}}\,\int_{-\omega}^{0}  d\omega \, 
\text{Tr}\left[\hat{\Gamma}_{\mathbf{k}^{\prime}} \hat{\mathcal{A}}(\mathbf{k}^{\prime},\omega) 
\hat{\Gamma}_{\mathbf{k}^{\prime}} \hat{\mathcal{A}}(\mathbf{k}^{\prime},\omega) \right] \nonumber 
\end{eqnarray}
where 
$(\hat{\Gamma}_{\mathbf{k}^{\prime}})_{\mathbf{n},\mathbf{n}^{\prime}}= \partial^2 \xi_{\mathbf{k}}/ \partial k_x \partial k_y \left|_{\mathbf{k}^{\prime}+ \mathbf{Q}_{\mathbf{n}}} \right. \, \delta_{\mathbf{n},\mathbf{n}^{\prime}}$
is a 16$\times$16 \BN Raman vertex matrix in the reduced Brillouin zone.

\clearpage

\bibliographystyle {apsrev4-1} 
\bibliography{cuprates}

\end{document}